\documentclass[twocolumn]{aa_aras} 

\usepackage{txfonts}
\usepackage{longtable}
\usepackage[left,modulo]{lineno}
\usepackage{hyperref}
\hypersetup{
  colorlinks=true,
  citecolor=blue,
  linkbordercolor={1 0 0},
  linkcolor=blue,
  urlcolor=blue,
  breaklinks=true
}
\usepackage{graphicx}
\usepackage{float}

\newcommand{\eps}{$\epsilon$ Aurigae\ }
\newcommand{\ha}{H$\alpha$\ }

\begin{document}
%
   \title{H$\alpha$ spectral monitoring of $\epsilon$ Aurigae\ 2009-2011 eclipse}


   \author{B. \textsc{Mauclaire}\inst{1}, C. \textsc{Buil}\inst{2}, T. \textsc{Garrel}\inst{3}, R. \textsc{Leadbeater}\inst{4}, A. \textsc{Lopez}\inst{5}
          }

   \institute{Observatoire du Val de l'Arc, 13530 Trets, France \\ \email{bma.ova@gmail.com} 
         \and Observatoire de Castanet, 31320 Castanet, France.
         \and Obsveratoire de Foncaude, 34990 Juvignac, France.
         \and Three Hills Observatory, The Birches, CA7 1JF, UK.
         \and Observatoire du Canet, 06110 Le Canet, France.
       }

\authorrunning{B. \textsc{Mauclaire} and al.}



   \date{Submitted to JAAVSO February 21, 2012. Accepted May 9, 2012. JAAVSO Volume 40, 2012.}

\abstract
{}
{We present and analyze \eps data concerning the evolution of the H$_\alpha$ line on the occasion of the 2009 International observation campaign launched to cover the eclipse of this object.}
{About 250 high resolution spectra of the \ha line were obtained by amateur covering the three years around eclipse. We visually inspect the dynamical spectrum constructed from the data and analyze the evolution with time of the EW (Equivalent Width) and of the radial velocity.}
{The spectroscopic data reveal many details which confirm the complexity of the \eps system. The object is far from being understood. In particular, according to our measurements, the eclipse duration has been underestimated. A complete analysis of details revealed by our data would require much time and effort. Observers are encouraged to continue monitoring the \ha line out of eclipse in the hope that it will provide further important information.
}
{}



   \maketitle
%

\begin{large}

\section{Introduction}

\eps is one of the most intriguing eclipsing star systems which has puzzled astronomers for nearly 200 years. The main eclipsing period is close to 27.1 years and the first spectroscopic surveys were undertaken during the 1929 and 1956 eclipses. A large campaign was also organized for 1982-1984. For a review of literature prior to the 2009-2011 eclipse, see \cite{guinan}. There are also numerous papers being prepared as a result of the 2009-2011 eclipse. Despite the concentrated efforts, some aspects of \eps remains a mystery. 

\eps is classified as an A8Iab star in an Algol type eclipsing binary system (\cite{simbad}). The prevailing model is of an F-type star with a hot clumpy Hydrogen disc and an object of unknown nature which produces an eclipse phenomenon lasting almost 2 years every 27.1 years. There may be a mid-eclipse brightening but solar proximity makes the photometry suspect at those times. Recently, it has been suggested that the eclipsing object is a 550~K dusty disk seen edge on, heated on the side facing the F star to 1100~K which may contain a B5V star (\cite{nl24}) that could contribute to emission wings surrounding \ha line. Light curves feature 0.1 magntude variations both inside outside eclipse. Variations have also been observed in the Equivalent Width (EW) and Radial velocity of spectral lines outside eclipse. These variations might be F star oscillations and wind.

During the 1982-1984 eclipse, this star was studied by amateur observers using multiband photometric methods. The \eps system was not clearly described despite all the acquired data.

Twenty-seven years later, an international campaign was organized to manage both spectroscopic and photometric observations by amateur observers with the aim of producing data with improved time resolution compared with that achieved during previous eclipses. In this article, we present amateur spectroscopic \ha line monitoring from February 12, 2008 to November 12, 2011.

\section{Observations}

In 2008, Jeff \textsc{Hopkins}\footnote{\url{http://www.hposoft.com/Campaign09.html}} organized the international observation compaign of the 2009 \eps eclipse. We acquired 247 high resolution spectra of the \ha line covering the three years around eclipse. These show significant variability throughout this period. The effect of the eclipse is clearly seen in this line from the end of April 2010 to end of April 2011. The spectra used in this study were recorded by five observers in Europe.

Most observations were made using LHIRES3\footnote{LHIRES3 and eShell are products from Shelyak Instruments, Grenoble, France: \url{http://www.shelyak.com}} spectrographs. C. Buil used an eShell spectrograph that covers wavelengths from 4500~\AA\ to 7000~\AA. Telescope diameters were between 0.2~m and 0.3~m. Spectral resolution is above 10,$\,$000 and most of the time around 15,$\,$000. Mean exposure time was 2,$\,$000~s. All setups are reported in Table \ref{table:1}.

\begin{table*}[!ht]
\caption{Equipment information for observation of $\epsilon$ Aurigae.}             
\label{table:1}      
\centering                          
\begin{tabular}{c c c c c c c}        
\hline\hline                 
Observer name & Telescope & Spectrograph & Resolution & Mean SNR & Spectral range ($\AA$) & Number of spectra \\    
\hline                        
   C. \textsc{Buil} & SCT 0.28 m & eShell & 10$\,$000 & 164 & 4500-7000 & 96 \\
   T. \textsc{Garrel} & SCT 0.21 m & Lhires3 2400 l/mm & 15$\,$000 & 108 & 6500-6630 & 87 \\
   B. \textsc{Mauclaire} & SCT 0.30 m & Lhires3 2400 l/mm & 15$\,$000 & 224 & 6520-6690 & 34 \\
   R. \textsc{Leadbeater} & SCT 0.25 m & Lhires3 2400 l/mm & 15$\,$000 & 98 & 6500-6700 & 25 \\
   A. \textsc{Lopez} & SCT 0.28 m & Lhires3 2400 l/mm & 15$\,$000 & 164 & 6500-6700 & 5 \\
\hline                                   
\end{tabular}
\end{table*}


\section{Reduction and analysis method}

Raw observations are available from Robin Leadbeater's \eps survey web page\footnote{\href{http://www.threehillsobservatory.co.uk/astro/epsaur\_campaign/epsaur\_campaign\_spectra\_table.htm}{\texttt{http://www.threehillsobservatory.co.uk/epsaur}}}. Spectra were reduced using standard procedures to produce calibrated and normalized lines profiles. Most of the reduction and analysis were done using pipelines and facilities available in SpcAudace\footnote{\url{http://spcaudace.free.fr}} package, part of Audela\footnote{\url{http://audela.org}} software. Reduction steps were: preprocessing, geometric corrections and registration. Then lines profiles were extracted with sky background subtraction. Wavelength calibration was done using calibration lamp spectra before and after each acquisition series to minimise the effects of calibration drifts. The instrumental response was then removed. An offset was then computed using telluric lines to achieve a final a wavelength calibration RMS uncertainty of 0.03 \AA. Finally heliocentric velocity correction was applied depending on the observation date. All wavelengths $\lambda$ are given in \AA\ (Angstr\"om).

\indent

Equivalent Width measurements were computed between $\lambda$6550 and $\lambda$6577 using linear integration and an extracted continuum obtained from a fit to the local star continuum\footnote{About EW's computation: \href{http://spcaudace.free.fr/docs/ew/}{\texttt{http://spcaudace.free.fr/docs}}}. The Chalabaev algorithm (\cite{chala83}) was used to estimate the uncertainty, which is mostly dependent on the signal-to-noise ratio, and appears to overestimate the uncertainty compared with the actual scatter observed around the long term trend.

\indent

Radial velocities are computed in two steps because the \ha line is asymmetric:
\begin{enumerate}
\item The Gaussian flank of the line was reproduced on the opposite side of the symmetry axis (theoretical wavelength of the line) and shifted to fit the line's opposite flank;
\item A Gaussian fit of these two flanks gave a measure of the line center.
\end{enumerate}


A dynamical spectrum was computed using 177 spectra corrected to heliocentric velocity and cropped to $\lambda$6550-$\lambda$6575. A linear interpolation was used to produce an image with a 1-day sampling interval. Such interpolation doesn't introduce bias for our analysis as the purpose of Figure \ref{dynagraph} is to show global behavior of the eclipse spread over several hundred days. Dates are logged in MJD (that is, JD$-$2400000). All computations were performed in SpcAudace. The monitoring covers a period of 719 days. Most of the information generated by our monitoring of the \ha line is contained in this dynamical spectrum. 

Analyzing this complex image turns out to be cumbersome, however. This is the reason why we have simplified the analysis by concentrating on the evolution with time of the EW that can be compared to V magnitude, and of the radial velocity that can be used to study eclipsing phenomena. Of course we have to keep in mind that EW loses its physical meaning when applied to complex line profiles lines, as it is the case for \eps , are likely to result from a combination of several sources. But before analyzing these quantities, let us first examine the spectral line profiles at dates that show important transitions.

\section{Behaviour of the wings}

Outside eclipse, the H$_\alpha$ line profile comprises a central absorption core flanked by emission features on the red and blue wings (see Fig. \ref{wings_before}). These features are highly variable as described by Golovin (\cite{golovin2008}).


   \begin{figure}[!ht]
     \centering
     \includegraphics[width=1.0\linewidth]{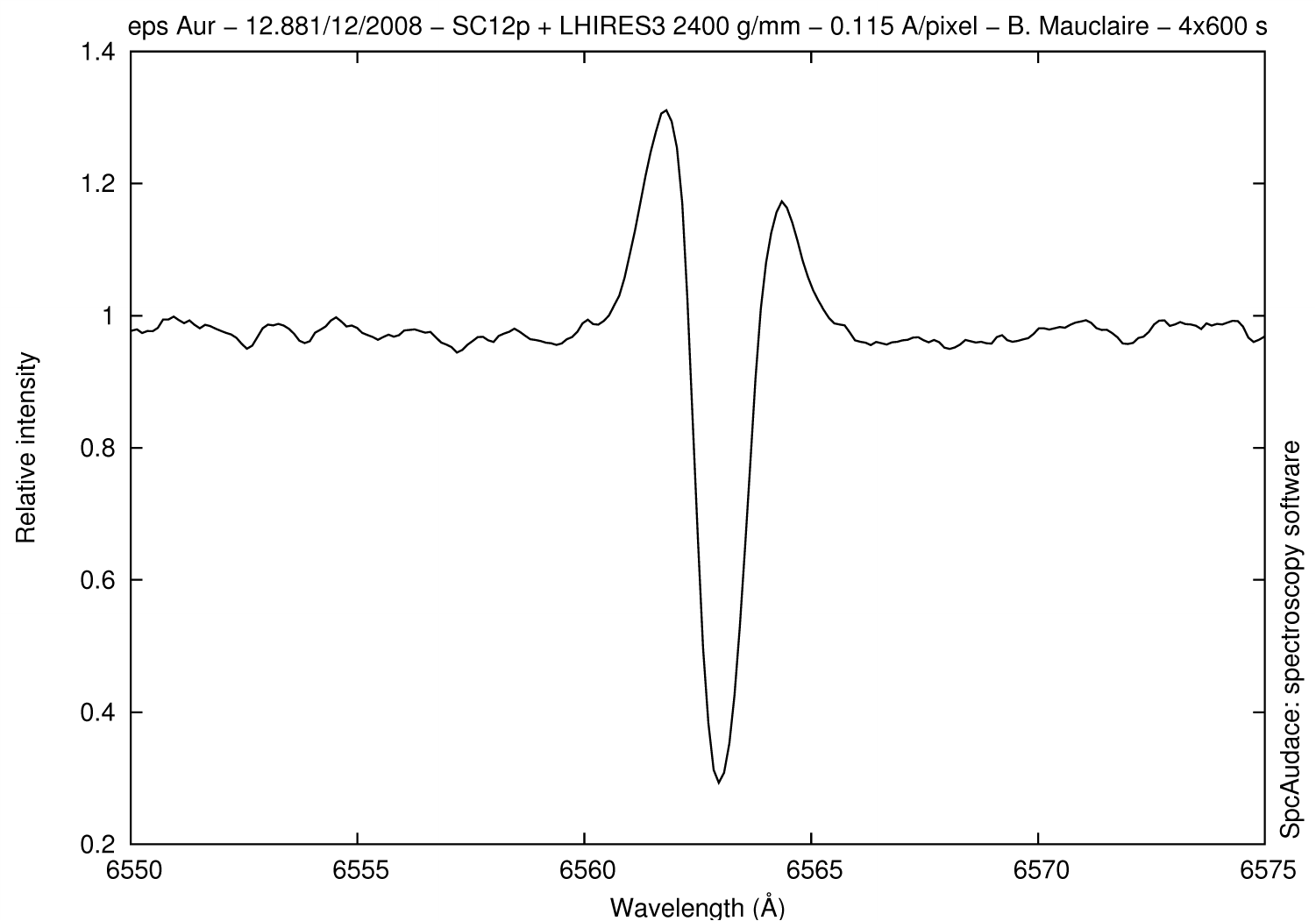}
     \caption{\eps spectrum showing \ha wings out of eclipse phase. Observations of 12.881/12/2008, SCT 0.3~m, Lhires3 2400 g/mm, 0.115 \AA/pixel, B. \textsc{Mauclaire}, 4$\times$600~s.}
     \label{wings_before}
   \end{figure}


In the region of the \ha line are absorption lines identified as telluric lines at $\lambda$6543.91, $\lambda$6547.71, $\lambda$6548.32, $\lambda$6552.63, $\lambda$6557.17, $\lambda$6568.81, $\lambda$6572.09, $\lambda$6574.85 and $\lambda$6586.68.

As we can see in Figures \ref{wings_evol} and \ref{dynagraph}, from MJD 55250 onwards there was additional absorption in the core which broadened rapidly, engulfing first the red emission and by MJD 55340 also the blue emission component. Note this is in contrast to the K\textsc{I} $\lambda$7699 line absorption which started decreasing in intensity during this phase (\cite{sas2011}).

   \begin{figure}[!ht]
     \centering
     \includegraphics[width=\linewidth]{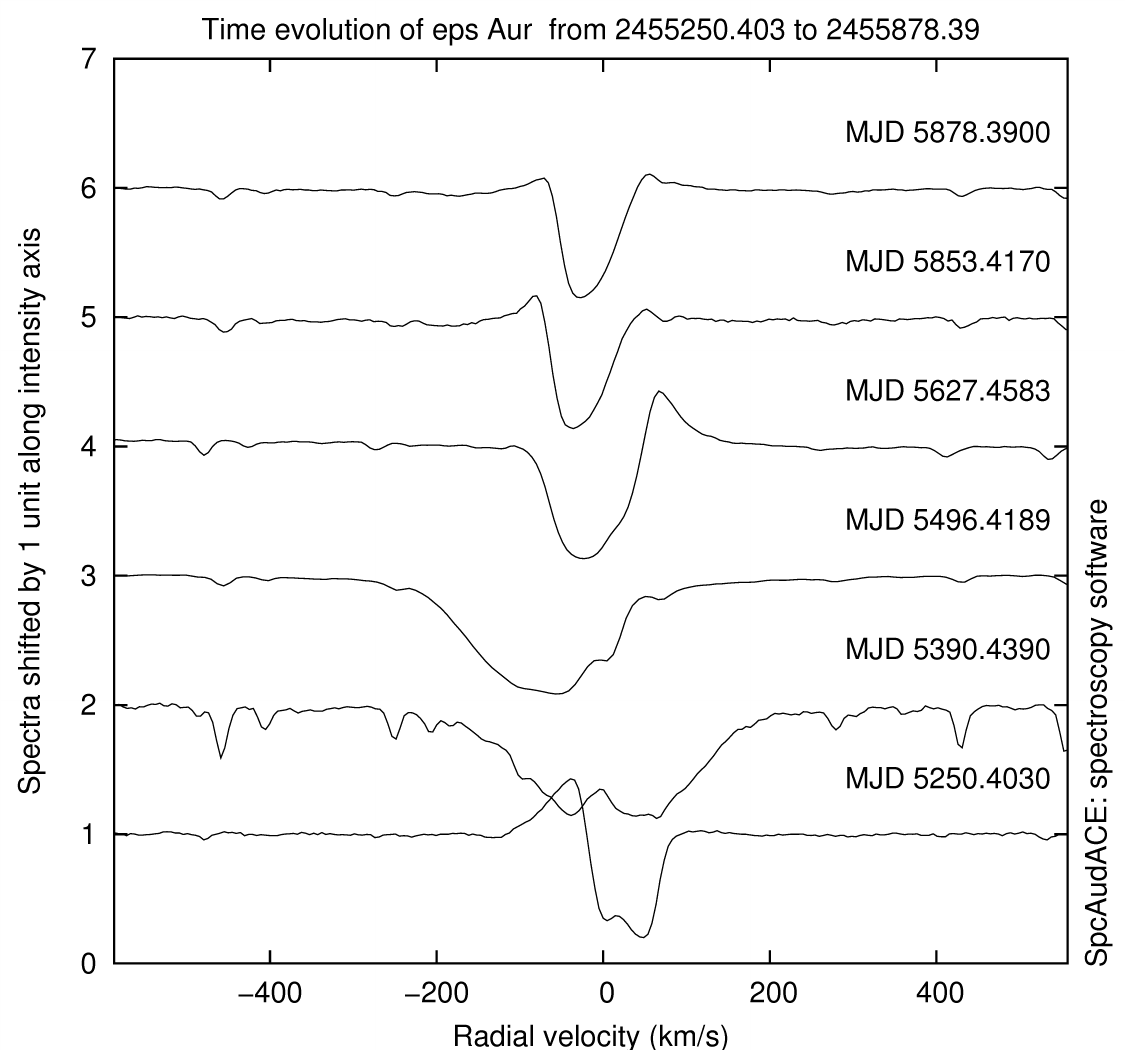}
     \caption{Time evolution of \eps from JD 2455250.403 to JD 2455878.39. \ha wings are shown at key dates.}
     \label{wings_evol}
   \end{figure}

During ingress and into totality (see spectra at MJD 55390.44 and MJD 55496.42) through the mid-eclipse point, the absorption core deepened and broadened slightly on the red side. The additional absorption moved to the blue and, at MJD 55520, the red emission feature reappeared.

At the beginning of the decreasing phase (see spectrum at MJD 55627.46), the \ha line became narrower, with an emission component at the red side. Then, from MJD 55853.42, the blue edge emission component returned as just before eclipse. After the main eclipse phase (see spectrum at MJD 55878.39), the blue and red wings were both present but small, starting to resemble the preeclipse profile (Figure~\ref{wings_before}).

At the end of the survey period there still appears to be an excess absorption on the blue side of the central absorption region compared with typical pre-eclipse spectra, possibly due to the continued presence of the eclipsing disc. However, the inherent variability of this at all phases makes the statement uncertain.

Today's understanding (\cite{stencelpriv}) is that the F star is semi-stable and capable of producing variability in lines in and out of eclipse. The disk is only modifying the optical spectrum during its passage.

\section{Equivalent width evolution with time}

EW measurements were computed between $\lambda$6550 and $\lambda$6577. Although the signal-to-noise ratio varies between observations and includes telluric lines which impact on EW, the effect most of the time is rather small. The data quality allows a reliable estimation of the EW. As mentioned earlier, the single quantity EW is a gross simplification of the complex nature of the line.

However, this quantity is the integral of the distribution of luminosity versus wavelength. It can thus be compared to similar integrals such as the V magnitude. As shown in Figures \ref{ew_evol} and \ref{magv_evol}, equivalent width (EW) and V magnitude (V mag.) are anti-correlated. Given that EW$>$0 for absorption lines, the eclipsing object occults the F star Hydrogen disk as first minimum in EW and in V mag. evolution are both close to MJD 55250, and as second minimum and V magnitude evolution are also both close to MJD 55630. These dates define totality inner limits (see Table \ref{table:2}).


\begin{table}[!ht]
\centering                          
\begin{tabular}{c| c}        
Main contact & MJD (day) \\
\hline                        
\hline
first & $55\,070\pm3$ \\
second & $55\,250\pm2$ \\
third & $55\,630\pm2$ \\
fourth & $55\,800\pm3$ \\
\hline                       
\end{tabular}
\caption{Contact dates during \eps eclipse.}             
\label{table:2}      
\end{table}

\vspace{-.5cm}

While second and third contacts (and mid-eclipse) times are well-defined by our EW as a function of Date trend (Figure~\ref{ew_evol}), the definition of first and fourth contact times are less obvious here and do not correspond to photometric contacts: these dates (second and third contacts) are likely to be linked with the densest ends of the eclipsing object.

   \begin{figure}[!ht]
     \centering
     \includegraphics[width=1.0\linewidth]{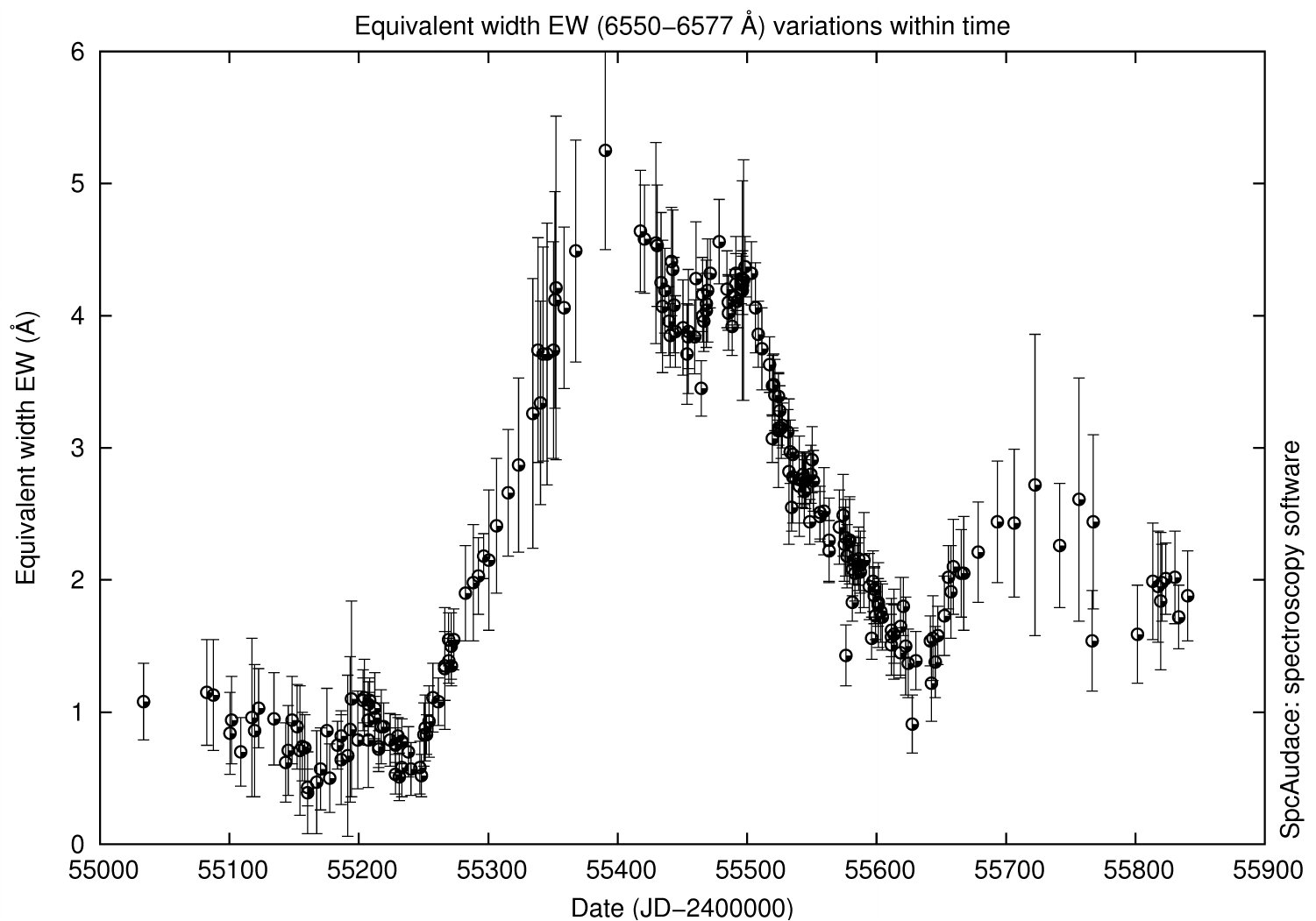}
     \caption{Plot showing evolution with time of Equivalent Width (EW) computed between $\lambda$6550 and $\lambda$6577. Note the two minima at MJD 55250$\pm$2 and MJD 55630$\pm$2 corresponding to second and third contacts dates.}
     \label{ew_evol}
   \end{figure}


   \begin{figure}[!ht]
     \centering
     \includegraphics[width=\linewidth]{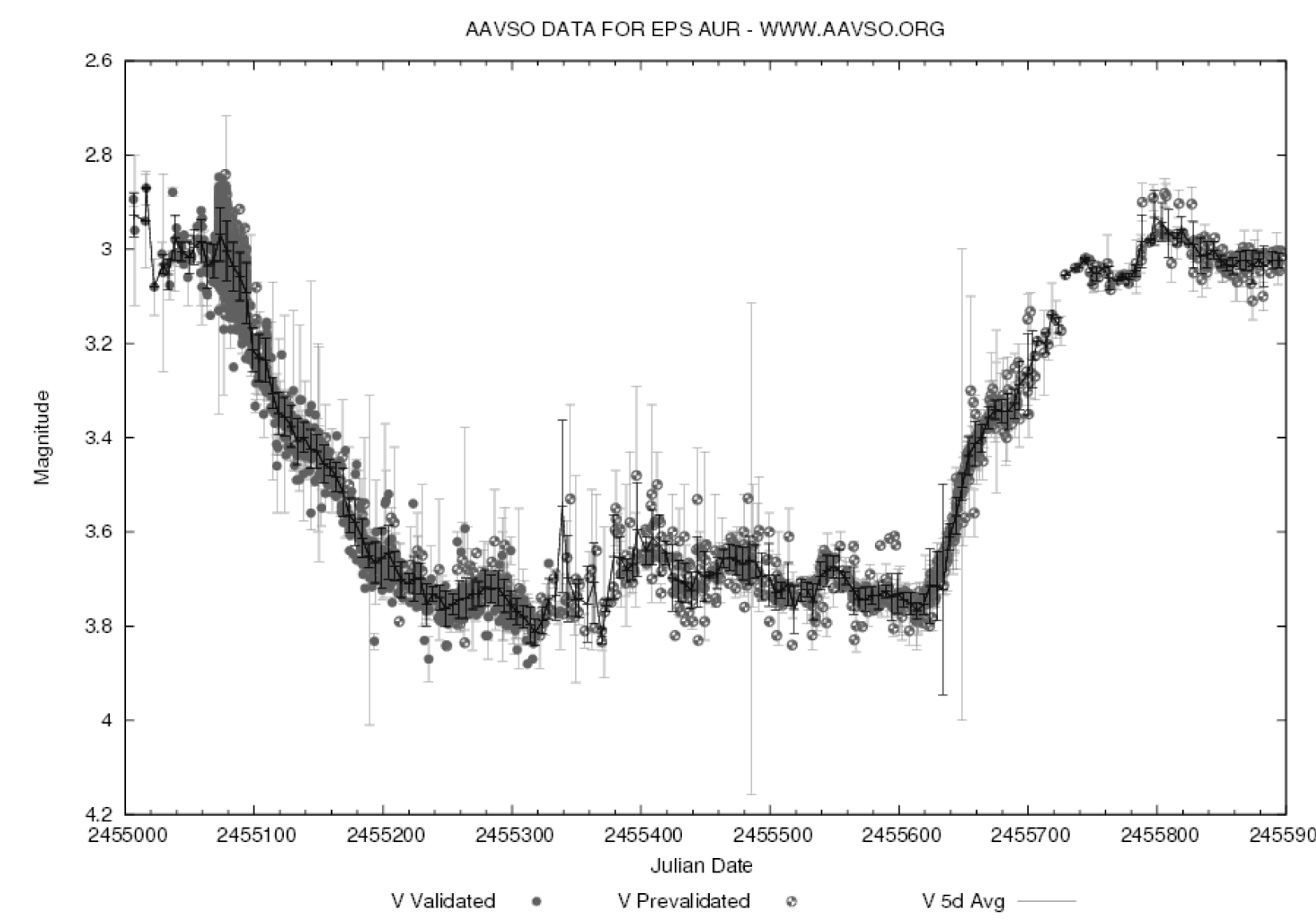}
     \caption{Light curve of \eps using data from the AAVSO International Database showing V magnitude evolution within time. Light curve courtesy of AAVSO.}
     \label{magv_evol}
   \end{figure}

During the eclipse phase and outside it too, there are many small variations in EW and V mag. This suggests that the occulting object and F star Hydrogen disk may be clumpy. The F star may have also an intrinsic pulsating activity (\cite{kemp85} and \cite{stencelpriv}) that produces such variations.

\ha EW has irregular variations like small steps during its increasing and decreasing phases. Similar behavior has been observed on the K\textsc{I} $\lambda$7699 asborption line. It has been interpreted as an indication of structures (possibly ring-like) within the disc (\cite{leadbeater2010}). Continued observation during egress may help to clarify this.

\section{Radial velocity evolution with time}

Figures \ref{wings_evol} and \ref{dynagraph} show how shapes are shifted in radial velocity. During eclipse (see spectra at MJD 55390.44 and MJD 55496.42), the absorption line became red shifted ($+14.79 \pm 1.37$~km/s). During the end phase of the eclipse (see spectrum at MJD 55627.46), the \ha line first returned to the position seen at MJD 55390.44 and then the absorption line became blue shifted ($-31.59 \pm 1.54$~km/s at MJD 55853.42). See Table~\ref{table:3} for measurements at key dates.

\begin{table}[!ht]
\centering                          
\begin{tabular}{c| c}        
MJD (day) & Radial velocity (km/s) \\
\hline                        
\hline
55$\,$199.24 & $+19.46\pm1.54$ \\
55$\,$390.44 & $+14.79\pm1.37$ \\
55$\,$436.49 & $-38.10\pm1.54$ \\
55$\,$521.46 & $-60.36\pm1.54$ \\
55$\,$819.46 & $-31.59\pm1.54$ \\
\hline                       
\end{tabular}
\caption{Radial velocity measurements of \eps at key dates.}             
\label{table:3}      
\end{table}

   \begin{figure}[!ht]
     \centering
     \includegraphics[width=\linewidth]{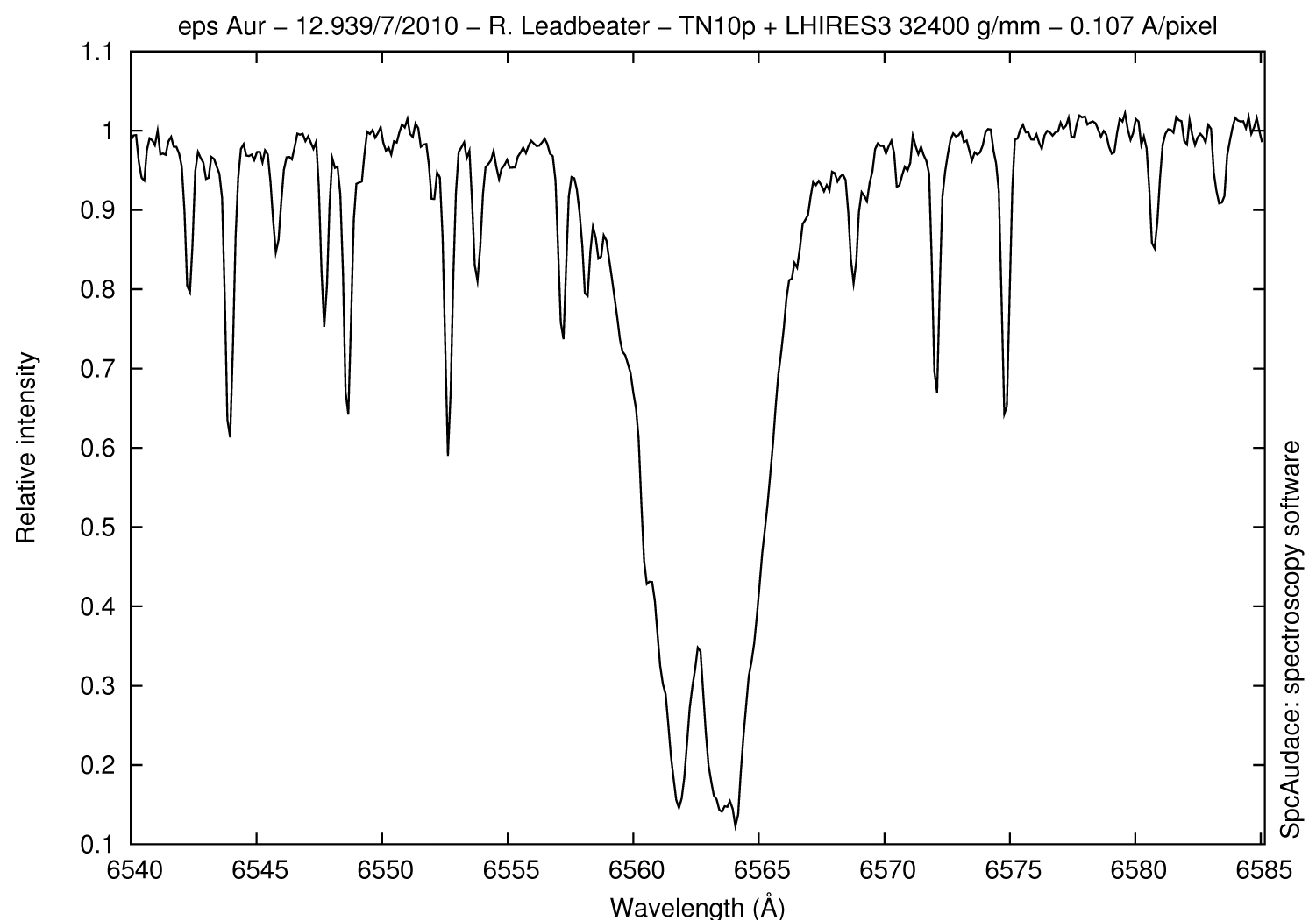}
     \caption{Spectrum of \eps showing emission component at bottom of \ha absorption line. Observations of 12.939/7/2010, R. \textsc{Leadbeater}, TN 0.25~m, Lhires3 2400 g/mm, 0.107 \AA/pixel.}
     \label{central_pic}
   \end{figure}

An emission component (Figure~\ref{central_pic}) appeared in the core of the \ha line close to the rest wavelength from MJD 55150 onward as the absorption increased in this region. This became more clearly defined as the surrounding flux level dropped further and moved across the region from red to blue. The shape of the emission component is revealed as the absorption region broadened and swept across it through mid eclipse. It is clear that the constant emission component is only revealed as the surrounding flux level drops.

   \begin{figure}[!ht]
     \centering

     \includegraphics[width=\linewidth]{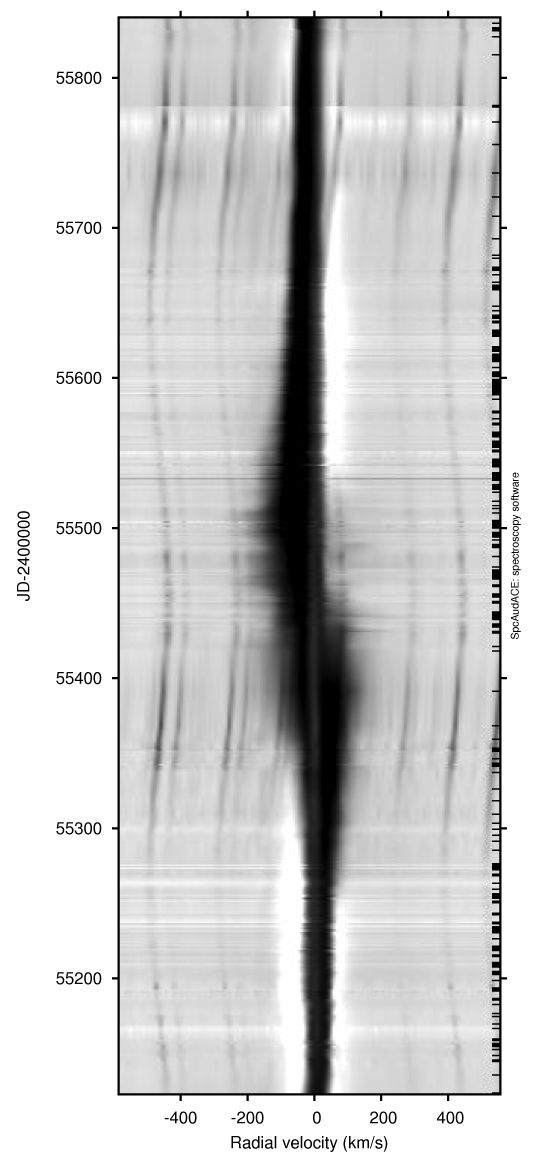}

     \caption{Dynamical spectrum of \ha line from JD 2455120.46 to JD 2455878.39. Interpolation between spectra was used to get a smooth image. Sinusoidal lines at both \ha line edges are telluric lines lying in line profiles that are corrected from heliocentric velocity. The black dashes along the right axis show the actual observation dates.}
     \label{dynagraph}
   \end{figure}

During our survey, the emission component measured by gaussian fitting remained centered on $6562.71~\pm~0.03$~\AA, which, in terms of radial velocity, amounts to $-4.97~\pm~1.54$~km/s. This does not account for the \eps systemic radial velocity estimated at $-2.26~\pm~0.15$~km/s (\cite{stefanik}), leading to a correction of $+0.049$~\AA\ (that is, $+2.26$~km/s). Note that Figure~\ref{dynagraph} is not well enough resolved to see such a small shift. Computations were done on the line profiles.

\section{In quest of new models}

The light variations of \eps in and out of eclipse have been the object of many studies.

During the 1983 eclipse \textsc{Kemp} and \textsc{Henson} (\cite{kemp85}) analyzed polarization data. They suggested that the F star is a non-radial pulsator and that its surrounding disk is tilted with respect to the orbit.

In 1991, \textsc{Ferluga} (\cite{ferluga1991}) suggested that to explain the shape of the light curve, the disk is not a continuous aggregate of dust, but instead a series of rings with a Cassini-like division. This model was more or less validated by observations.

During this eclipse, a wide variety of observations have been undertaken: infrared, ultraviolet, interferometry, photometry, and high resolution spectral monitoring. Thus, a considerable amount of information is now available (see \cite{stencel2010} and \cite{nl24} for an overview). It now remains to develop a model that fits all the data at hand. Undoubtedly, the high resolution spectral monitoring data will be very important for constraining theses models.

\section{Conclusions}

Our Ha monitoring of \eps shows that, contrary to what was forecast, the effects of the eclipse extended beyond December 2011. Post-eclipse observations are needed. R. \textsc{Stencel} welcomes any outside eclipse spectroscopic contributions to the campaign over the coming months and years, especially those covering the Na D lines.

We have observed similarities and discrepancies between the EW and V magnitude evolution with time. The discrepancies remain to be explained, but that is beyond the scope of this article. We also were able to define key dates in the eclipsing phenomenon. However, much remains to be analyzed. Obviously the \ha monitoring brings a lot of information which should place many constraints on the models conceived by scientists about $\epsilon$ Aurigae.

Amateur spectroscopists are now able to monitor bright targets with a spectral resolution of about 15,$\,$000. Suitably equipped amateurs constitute a team with long term monitoring capacity which is widely distributed over the planet.

In any case, we hope that this information will help scientists to solve the mysteries hidden behind this fascinating object.

\bigskip
\begin{acknowledgements}
This campaign would not have been possible without Jeff \textsc{Hopkins}'s motivation and constant efforts to encourage amateur observations. The authors acknowledge Robert \textsc{Stencel} for his help and encouragements thoughout this campaign. They also kindly thank the referee for constructive comments and very helpful work.

\noindent All of the data presented in this paper were obtained from amateur observers. The authors also acknowledge Eric \textsc{Barbotin}, St\'ephane \textsc{Charbonnel}, Val\'erie \textsc{Desnoux}, Stanley \textsc{Gorodenski}, Keith \textsc{Graham}, Torsten \textsc{Hansen}, James \textsc{Edlin}, Brian E. \textsc{McCandless}, \'Eric \textsc{Sarrazin}, Lothar \textsc{Schanne}, Jos\'e \textsc{Ribeiro}, Fran\c{c}ois \textsc{Teyssier}, Olivier \textsc{Thizy} and John \textsc{Strachan} for all the work they carried out at other wavelengths and spectral resolutions on this mysterious star.
\end{acknowledgements}



\end{large}
\end{document}